\documentclass[aps,prl,twocolumn,showpacs,superscriptaddress,showkeys]{revtex4-1}

\usepackage{amssymb}
\usepackage{amsmath}
\usepackage{amsfonts}
\usepackage{graphicx}
\usepackage{color}
\usepackage{xspace}
\usepackage{ulem}
\usepackage{mathtools}
\usepackage{hhline}

\newcommand{\AddrUG}{Departamento de F\'{\i}sica Te\'orica y del Cosmos, Universidad de Granada, Granada-18071, Spain}
\newcommand{\AddrEdin}{School of Physics and Astronomy, University of Edinburgh, Edinburgh, EH9 3FD, United Kingdom}
\newcommand{\AddrUERJ}{Departamento de F\'isica Te\'orica, Universidade do Estado do Rio de Janeiro, 20550-013 Rio de Janeiro, RJ, Brazil}
\newcommand{\AddrAve}{Departamento de F\'{\i}sica da Universidade de Aveiro and CIDMA,  Campus de Santiago, 3810-183 Aveiro, Portugal}

\begin{document}


\title{Warm Little Inflaton}

\author{Mar Bastero-Gil}    \email{mbg@ugr.es}\affiliation{\AddrUG}
\author{Arjun Berera}    \email{ab@ph.ed.ac.uk}\affiliation{\AddrEdin} 
\author{Rudnei O. Ramos}    \email{rudnei@uerj.br}\affiliation{\AddrUERJ}
\author{Jo\~{a}o G.~Rosa} \email{joao.rosa@ua.pt\\ Also at Departamento de F\'{\i}sica e Astronomia, Faculdade de Ci\^encias da Universidade do Porto, Rua do Campo Alegre 687, 4169-007 Porto, Portugal.}\affiliation{\AddrAve}

\date{\today}

\begin{abstract}
We show that inflation can naturally occur at a finite temperature $T>H$ that is sustained by dissipative effects, when the inflaton field corresponds to a pseudo Nambu-Goldstone boson of a broken gauge symmetry. Similarly to ÒLittle HiggsÓ scenarios for electroweak symmetry breaking, the flatness of the inflaton potential is protected against both quadratic divergences and the leading thermal corrections. We show that, nevertheless, nonlocal dissipative effects are naturally present and are able to sustain a nearly thermal bath of light particles despite the accelerated expansion of the Universe. As an example, we discuss the dynamics of chaotic warm inflation with a quartic potential and show that the associated observational predictions are in very good agreement with the latest Planck results. This model constitutes the first realization of warm inflation
requiring only a small number of fields; in particular, the inflaton is directly coupled to just two light fields.
\end{abstract}


\maketitle



Alongside its success in explaining the present flatness and homogeneity of our Universe, inflation  \cite{inflation} may provide one of the best probes of high-energy fundamental physics. A key goal in modern cosmology is thus to incorporate the physical mechanism driving inflation, presumably associated with a new fundamental scalar field, into a more complete particle physics framework.

Most of the recent literature has focused on finding, within extensions of the Standard Model, flat potentials that can sustain a slowly evolving scalar field, which mimics a cosmological constant, for 50$-$60 {\it e}-folds of inflationary expansion. Although this is a necessary and important task, it discards other potentially important effects of interactions between the new inflaton field and other particle degrees of freedom that may play a crucial role in embedding inflation within a larger framework.

One of these effects is nonequilibrium dissipation, which results from the energy exchange between the inflaton field and other quantum fields in the cosmic plasma. In the leading adiabatic approximation, it is well known that this leads to an additional friction term in the inflaton equation of motion  \cite{Berera:1995wh,Berera:1995ie}:
\begin{eqnarray} \label{inflaton_eq}
\ddot\phi +3H\dot\phi + \Upsilon\dot\phi+ V'(\phi) = 0~,
\end{eqnarray}
where dots correspond to time derivatives, primes denote here derivatives with respect to $\phi$ and $H$ is the Hubble parameter. Such a friction term is often thought to have no significant effect during the slow roll phase, since {\it a priori} accelerated expansion  quickly dilutes the cosmic plasma. However, by transferring the inflaton's energy into the plasma, dissipation provides a source that can compensate for this effect. In particular, when dissipation results from interactions with light degrees of freedom (DOF) that thermalize within a Hubble time, the evolution of the radiation energy density is given by:
\begin{eqnarray} \label{radiation_eq}
\dot\rho_R+4H\rho_R= \Upsilon\dot\phi^2~,
\end{eqnarray}
where $\rho_R= (\pi^2/30) g_* T^4$ for $g_*$ relativistic DOF~at temperature $T$. In the slow roll regime the source term varies adiabatically, with $\Upsilon=\Upsilon(T, \phi)$ in general, and the radiation fluid may reach a slowly evolving state where $\rho_R\simeq \Upsilon\dot\phi^2/4H$. This may sustain a temperature $T\gtrsim H$ during inflation for $\dot\phi \gg H^2$, even for $\Upsilon < H$, without violating the slow roll condition that $\dot\phi \ll \sqrt{V(\phi)}\simeq H M_P$, where $M_P$ denotes the reduced Planck mass.

The presence of dissipative effects may thus lead to a {\it warm} rather than {\it supercooled} inflationary regime, an observation that was first made more than two decades ago \cite{Berera:1995ie}. This idea has several attractive features, namely that the additional friction may alleviate the required flatness of the potential. The slow roll conditions are, in particular, modified in the presence of dissipation to $\epsilon_\phi, |\eta_\phi| < 1+Q$, where $Q=\Upsilon/3H$ and $\epsilon_\phi= M_P^2(V'/V)^2/2$ and $\eta_\phi = M_P^2 V''/V$ are the slow roll parameters \cite{Berera:2008ar, BasteroGil:2009ec}. Moreover, in the slow roll regime, one can show that
\begin{eqnarray} \label{radiation_eq}
{\rho_R\over V(\phi)}\simeq {1\over 2}{\epsilon_\phi\over 1+Q}{Q\over 1+Q}~,
\end{eqnarray}
so that radiation, although subleading during inflation (as required for accelerated expansion), may smoothly become the dominant component if $Q\gtrsim 1$ when $\epsilon_\phi\sim 1+Q$, with no need for a separate reheating period \cite{Berera:1996fm}. In addition, dissipation modifies the growth of inflaton fluctuations \cite{Berera:1999ws, Hall:2003zp, Moss:2007cv, Graham:2009bf, Ramos:2013nsa}, leaving a distinctive imprint on the primordial spectrum that can be used to probe the interactions between the inflaton and other particles.

It was realized a few years after its original proposal \cite{BGR, YL}, however, that the idea of warm inflation was not easy to realize in concrete models, with Ref.~\cite{YL} going further to suggest it could simply not be possible. First, it is hard to couple the inflaton directly with light fields. Considering, e.g.,~a Yukawa interaction $g\phi\bar\psi\psi$, the fermion acquires a mass  $m_\psi=g\phi$ that is large unless the coupling is very suppressed, taking into account the large inflaton values typically required by the slow roll conditions. A small coupling then implies that dissipative effects may be too small to sustain a thermal bath at temperature $T>H$. Second, a direct coupling to light fields may lead to large thermal corrections to the inflaton mass $m_\phi \sim g T$, which could prevent slow roll for $T>H$. 

Successful models of warm inflation have nevertheless been found when the inflaton is only indirectly coupled to light DOF~through heavy mediator fields   \cite{Berera:2002sp, Moss:2006gt, BasteroGil:2010pb, BasteroGil:2012cm}. Thermal mass corrections are exponentially suppressed in this regime, whereas the dissipation coefficient is only suppressed by powers of $T/M_{m}\lesssim 1$, where $M_{m}$ is the mediator mass. This suppression implies, however, that a large multiplicity of mediator fields is required to sustain the thermal bath for 50$-$60 {\it e}-folds of inflation and, although technically consistent, this would mean that warm inflation can be realized only in special scenarios, such as, e.g., the case of the brane constructions discussed in Ref.~\cite{BasteroGil:2011mr}. 

In this Letter, we show, for the first time, that warm inflation can be realized by directly coupling the inflaton to a few light fields. Our scenario borrows some of the ingredients used in ``Little Higgs" models of electroweak symmetry breaking \cite{ArkaniHamed:2001nc}, where the Higgs boson is a pseudo-Nambu-Goldstone boson (PNGB) of a broken gauge symmetry and its mass is naturally protected against large radiative corrections (see \cite{Schmaltz:2005ky} for a review). In the same spirit, we take the inflaton to be a PNGB of a broken U(1) gauge symmetry, as considered in Refs.~\cite{Kaplan:2003aj, ArkaniHamed:2003mz}. 

The main idea is quite simple. Suppose that there are two complex Higgs fields, $\phi_1$ and $\phi_2$, with identical U(1) charges $q$ and that the scalar potential is such that both fields have a nonzero vacuum expectation value, which we take to be equal for simplicity, $\langle \phi_1\rangle= \langle \phi_2\rangle \equiv M/\sqrt{2}$. The phases of both fields then yield two NG bosons, but only one linear combination is the true NG boson that becomes the longitudinal component of the massive U(1) gauge boson upon symmetry breaking. The relative phase of the two fields is, on the other hand, a singlet, since U(1) transformations shift the phase of each field by the same amount. This scalar singlet thus remains as a physical DOF in the broken phase. It is convenient to parametrize the fields in the broken phase in the form
\begin{eqnarray}
\phi_1 = {M\over\sqrt{2}} e^{i\phi/M}~,\qquad \phi_2 = {M\over\sqrt{2}} e^{-i\phi/M}~, 
\end{eqnarray}
where we assume the radial Higgs fields to decouple for $T\lesssim M$. We thus take the inflaton to be the PNGB $\phi$, which being a gauge singlet may have an arbitrary scalar potential that can be sufficiently flat to sustain inflation.

We consider, in addition, that the Higgs fields are coupled to left-handed fermions $\psi_{1L}$ and $\psi_{2L}$ with U(1) charge $q$ as well as their right-handed counterparts $\psi_{1R}$ and $\psi_{2R}$, which we take to be gauge singlets. We consider identical couplings in magnitude and impose the interchange symmetry $\phi_1\leftrightarrow i\phi_2$, $\psi_{1L,R}\leftrightarrow \psi_{2L,R}$, such that the allowed Yukawa interactions are of the form
\begin{eqnarray} \label{Yukawa_inflaton}
-\mathcal{L}_{\phi \psi} &=& {g\over \sqrt{2}}(\phi_1+\phi_2)\bar\psi_{1L} \psi_{1R} - i{g\over \sqrt{2}}(\phi_1-\phi_2)\bar\psi_{2L} \psi_{2R}\nonumber\\
&=& gM\cos(\phi/M) \bar{\psi}_1\psi_1 + gM\sin(\phi/M) \bar{\psi}_2\psi_2~.
\end{eqnarray} 
The resulting Dirac fermion masses are thus  $m_{1,2}\leq gM$, such that they may remain light during inflation for an arbitrary inflaton value, provided that $gM\lesssim T \lesssim M$.

For $m_i\ll T$, $i=1,2$, the fermion contribution to the finite temperature effective potential is given by \cite{kapusta, Cline:1996mga}:
\begin{eqnarray} \label{finite_T_potential}
V_{Ti} \simeq -{7\pi^2\over180}T^4+{m_i^2T^2\over 12}+ {m_i^4\over 16\pi^2}\left[\ln\left({\mu^2\over T^2}\right)-c_f\right]~,
\end{eqnarray} 
where $\mu$ is the $\overline{\mathrm{MS}}$ renormalization scale and $c_f\simeq 2.635$. From here it is clear that, adding the contribution of both fermions, the quadratic term becomes independent of $\phi$, such that the leading thermal inflaton mass corrections cancel, leaving only the subleading Coleman-Weinberg term.
Analogously, one can expand the Yukawa interactions about the background inflaton value:
\begin{eqnarray}
\mathcal{L}_{\phi\psi}= -\sum_{i=1,2}\left(m_i + g_i \delta\phi + {f_i\over 2}\delta\phi^2 +\ldots \right)\bar{\psi}_i\psi_i~,
\end{eqnarray} 
to compute the inflaton self-energy. At one-loop order, the relevant diagrams are shown in Fig.~1.

\begin{figure}[htbp]
\centering\includegraphics[scale=0.5]{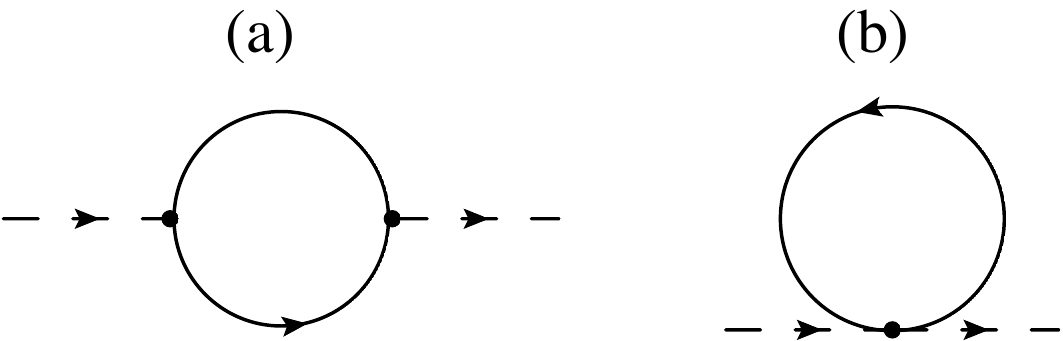}
\caption{Feynman diagrams contributing to the inflaton self-energy at one-loop order.}
\label{self-energy}
\end{figure}

It is then easy to show that, for $m_i \ll T$,  the zero-momentum self-energy is given by (see, e.g., \cite{kapusta})
\begin{eqnarray}
\Sigma_{\phi}(0)&=& \left[ \left(g_1^2 +m_1 f_1\right)+\left(g_2^2+m_2 f_2\right)\right] I_T\nonumber\\
&=& g^2\left[-\cos(2\phi/M) +\cos(2\phi/M)\right] I_T  =  0~,
\end{eqnarray} 
where the loop integral $I_T \simeq -(\Lambda^2/ 2\pi^2) + (T^2/ 6)$ to leading order for a momentum cutoff $\Lambda$. Hence, both quadratic divergences and thermal mass corrections cancel between the contributions of both fermions, which is typical of all Little Higgs models \cite{Espinosa:2004pn}.  Note that no other light fields contribute to the inflaton self-energy in our construction, since the inflaton is, in particular, a singlet field.

This cancellation occurs, however, only at the local level of the effective potential, whereas the dissipative term in Eq.~(\ref{inflaton_eq}) is the leading nonlocal correction to the effective action in the adiabatic approximation, where the inflaton's motion is slower than the relevant microphysical processes. It is given in terms of the retarded inflaton self-energy in the real-time formalism \cite{Berera:2008ar}
\begin{equation}
\Upsilon = \int d^4x' \Sigma_R(x,x') \, (t'-t)~.
\label{defdiss}
\end{equation}
At one-loop order, only Fig.~\ref{self-energy}(a) yields a nonlocal contribution, with external legs corresponding to different times $t$ and $t'$, whereas Fig.~\ref{self-energy}(b) contributes only locally. Equivalently, only Fig.~\ref{self-energy}(a) can be consistently ``cut" and contribute to the inflaton's decay. In this scenario, we thus eliminate the troublesome thermal corrections without suppressing dissipative effects. 

The retarded self-energy can be computed using standard techniques \cite{kapusta}, and here we only outline the main steps of the calculation, leaving the details for a companion paper. For $T\gg m_i$, the leading contributions correspond to on-shell fermions, for which \cite{BasteroGil:2010pb}
\begin{equation} \label{Ups2}
\Upsilon_i = 4 \frac{g_i^2}{T} \int \frac{d^3 p}{(2 \pi)^3} \frac{m_i^2}{\Gamma_{\psi_i} \omega_p^2}n_F(\omega_p) \left[ 1- n_F(\omega_p) \right]~,
\end{equation}
where $n_F(\omega_p)$ is the Fermi-Dirac distribution, $\omega_p=\sqrt{|\mathbf{p}|^2+m_i^2}$, and  $\Gamma_{\psi_i}$ is the fermion decay width. We assume that the latter is dominated by additional Yukawa interactions, involving a scalar singlet $\sigma$ and chiral fermions $\psi_{\sigma R}$, with charge $q$, and $\psi_{\sigma L}$, with zero charge:
\begin{eqnarray}
\mathcal{L}_{\psi\sigma}= -h\sigma \sum_{i=1,2}\left( \bar{\psi}_{iL}\psi_{\sigma R}+ \bar{\psi}_{\sigma L}\psi_{iR}\right)~,
\end{eqnarray} 
which respects the interchange symmetry. This yields for the on-shell decay width at finite temperature, neglecting the masses of the decay products,
\begin{equation} \label{decayGamma}
{\Gamma}_{\psi_i}= 
{h^2\over 16\pi }{T^2 m_i^2 \over \omega_p^2 |{\bf p}|}\left[F\left({k_+\over T}, {\omega_p\over T}\right)-F\left({k_-\over T}, {\omega_p\over T}\right)\right]~,
\end{equation}
where $k_\pm = (\omega_p \pm |\mathbf{p}|)/2$ and 
\begin{eqnarray}
F(x,y)&=& xy - \frac{x^2}{2} + (y-x) \ln \left(
\frac{1-e^{-x}}{1+e^{-y+x}} \right)  
\nonumber \\
&+& {\rm Li}_2\left(e^{-x} \right) + {\rm Li}_2 \left(-e^{-y+x}\right)~,
\end{eqnarray}
where ${\rm Li_2}(z)$ is the dilogarithm function. These Yukawa terms also give thermal corrections to the fermion masses, $\Delta m_i^2 \simeq  h^2T^2/8$, which dominate over the inflaton contribution for $h\gg g$ and $T\lesssim M$. Adding the contributions of $\psi_1$ and $\psi_2$ to dissipation, we then get
\begin{eqnarray}
\Upsilon = C_T T~, \qquad C_T\simeq \alpha (h) g^2/h^2~.
\end{eqnarray}
An approximate form of the numerical factor $\alpha(h)$ can be obtained by evaluating the fermion decay width at the momentum $p_{\mathrm{max}}\simeq 3.24 T$ that yields the largest contribution to the dissipation coefficient (\ref{Ups2}), yielding $\alpha(h)\simeq 3/[ 1-0.34\log (h)]$ \cite{note1}. 

For this dissipation coefficient, $Q=\Upsilon/3H\propto T/H$, and we may write the coupled inflaton and radiation equations in the slow roll regime in the form
\begin{eqnarray} \label{slow_roll_eqs}
{Q'\over Q}={6\epsilon_\phi-2\eta_\phi\over 3+5Q}~,\qquad {\phi'\over M_P}=-{\sqrt{2\epsilon_\phi}\over 1+Q}~,
\end{eqnarray} 
where primes denote derivatives with respect to the number of {\it e}-folds of inflation. Thus, the ratios $Q$ and $T/H$ grow during inflation for potentials with $6\epsilon_\phi-2\eta_\phi>0$, i.e., those yielding a red-tilted spectrum in supercooled scenarios.

Dissipation modifies the curvature power spectrum in different ways. First, it directly sources inflaton fluctuations, yielding a Langevin equation that generalizes (\ref{inflaton_eq}) for an inhomogeneous field. Inflaton particles may also be thermally excited, having a Bose-Einstein rather than vacuum phase space distribution. Finally, inflaton and radiation fluctuations are coupled due to the $T$ dependence of the dissipation coefficient. The resulting dimensionless power spectrum has the form  \cite{Berera:1999ws, Hall:2003zp, Moss:2007cv, Graham:2009bf, BasteroGil:2011xd, Ramos:2013nsa}
\begin{eqnarray} \label{spectrum}
\!\!\Delta_\mathcal{R}^2 =\frac{V_*(1 \!+\!Q_*)^2 }{24\pi^2 M_P^4\epsilon_{\phi_*}}\!\!\left(\!1\!+\!2n_* \!+\!\frac{2\sqrt{3}\pi Q_*}{\sqrt{3\!+\!4\pi Q_*}}{T_*\over H_*}\right)\! G(Q_*)
\end{eqnarray}
where $n_*$ denotes the inflaton phase space distribution and all quantities are evaluated when the relevant cosmic microwave background (CMB) modes become super-horizon 50$-$60 {\it e}-folds before inflation ends. The function $G(Q_*)$ accounts for the growth of inflaton fluctuations due to the coupling to radiation and must be determined numerically. We have extended the analysis in Ref.~\cite{BasteroGil:2011xd} and obtained the numerical fit:
\begin{eqnarray} \label{growing_mode}
G(Q_*)\simeq 1+ 0.0185Q_*^{2.315}+ 0.335 Q_*^{1.364}~.
\end{eqnarray} 
Tensor modes are essentially unaffected by the dissipative dynamics, due to the smallness of gravitational interactions, such that the enhancement of scalar curvature perturbations generically results in a decrease of the tensor-to-scalar ratio $r=\Delta_t^2/\Delta_\mathcal{R}^2$. It also results in a modified consistency relation between $r$ and the tensor index $n_t$ that may be used to distinguish warm and supercooled inflationary models and, thus, probe the interactions between the inflaton and other fields, as noted in Ref.~\cite{Bartrum:2013fia}.

Since the modifications to the curvature power spectrum depend only on $Q$ and $T/H\propto Q$, the scalar spectral index $n_s-1 \simeq d\ln\Delta_\mathcal{R}^2/d N_e$ will be determined, according to Eq.~(\ref{slow_roll_eqs}), by the combination $-6\epsilon_\phi+2\eta_\phi$ at horizon crossing, as in cold inflation. For instance, with thermalized inflaton fluctuations, $n_*\simeq T_*/H_*\gtrsim 1$, and $Q_*\ll1 $, we find $n_s \simeq 1 + (2/3)(2\eta_\phi-6\epsilon_\phi)$. In this case, we can also use the slow roll equations to show that $(T_*/H_*)^2\simeq 30 C_T/ (8\pi^4 \Delta_\mathcal{R}^2 g_*)\sim 10^6 C_T$, 
%
%
with $g_*\simeq 12.5$ in our model. This shows that a warm regime, $T\gtrsim H$, can be obtained for the entire duration of inflation if the couplings $g/h\gtrsim 10^{-3}$ and that $Q_*\gtrsim 10^{-7}$.

In general, one can use the amplitude of the curvature power spectrum (\ref{spectrum}) to determine $T_*/H_*$ and $Q_*$ for given values of the couplings, independently of the inflaton potential. We find that observationally consistent scenarios have $C_T\lesssim 0.02$, requiring no large couplings.

As an example, let us consider the case of chaotic inflation with a quartic potential: $V(\phi)=\lambda \phi^4$ \cite{note2}. The slow roll equations (\ref{slow_roll_eqs}) can be analytically integrated, and, e.g., for $Q_*\ll 1$ this yields for the number of {\it e}-folds
\begin{eqnarray}
N_e \simeq {1\over 8}\left({\phi_*\over M_P^2}\right)^2\left(1+1.1Q_*-Q_*\log Q_*\right)~,\
\end{eqnarray} 
 giving the leading correction to the supercooled result. To leading order, we find  $n_s-1\simeq -2/N_e$ for thermal fluctuations, which gives $n_s\simeq$0.96$-$0.967 for $N_e=$50$-$60, in agreement with the Planck results \cite{Ade:2015lrj}. 

\begin{figure}[htbp]
\centering\includegraphics[scale=1.15]{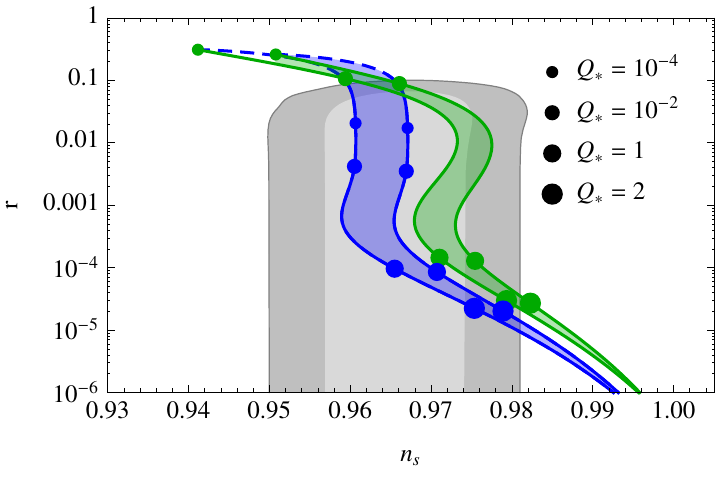} \vspace{-0.4cm}
\caption{Observables for the quartic model with 50$-$60 {\it e}-folds of inflation, considering nearly thermal (blue) and negligible (green) inflaton occupation numbers. The Planck 2015 68\% and 95\% C.L.~contours are shown in gray \cite{Ade:2015lrj}. The dashed curves, corresponding to $T_*< H_*$, are shown for completeness, although our analysis is not valid in this regime.}
\label{quartic_observables}
\end{figure}

In Fig.~\ref{quartic_observables}, we show the predictions for the quartic model for both the thermalized case and when inflaton particle production in the thermal bath is negligible, $n_*\ll 1$.

The ratios $Q_*$ and $T_*/H_*$ increase from top to bottom in Fig.~\ref{quartic_observables}, where it is clear that the tensor-to-scalar ratio is suppressed compared to the cold case. The regions within the Planck contours have, in both cases, $T_*/H_*\gtrsim 2$ and $Q_*\lesssim 1$, whereas if dissipation is already strong at horizon crossing, $Q_*\gtrsim1 $, the spectrum becomes more blue-tilted due to the coupling between inflaton and radiation fluctuations. Chaotic warm inflation is nevertheless consistent with observations down to $r\gtrsim 10^{-5}$, as opposed to the supercooled case \cite{Ade:2015lrj}. This agreement is obtained within consistent models for perturbative couplings $h=\mathcal{O}(1)$ and $g\simeq $ 0.05$-$0.2, with maximum temperature (at horizon crossing) $T_*=($1.3$-$6.7$)\times 10^{15}$ GeV, showing that CMB data can give precise information about the interactions between the inflaton and other fields and also the temperature during inflation.

\begin{figure}[htbp]
\centering\includegraphics[scale=1.15]{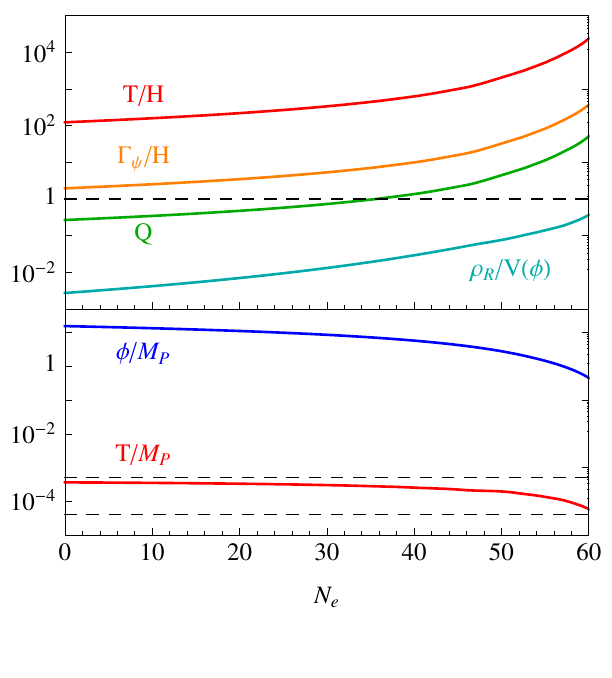}\vspace{-0.9cm}
\caption{Example of the dynamical evolution in warm inflation with a quartic potential. The dashed lines in the bottom plot correspond to $M$ and $gM$ (in Planck units).}
\label{quartic_dynamics}
\end{figure}

In Fig.~\ref{quartic_dynamics}, we illustrate the evolution of the different dynamical quantities in the quartic model, obtained numerically for an example with $g=0.08$, and $h=2$, yielding $T_*/H_*\simeq 123$ and $Q_*\simeq 0.27$. For $\phi_*\simeq 16 M_P$, the slow roll conditions fail after $\simeq 60$ {\it e}-folds, whereas in the absence of dissipation one would only get half this value. It is also clear that inflation ends in the strong dissipation regime and that radiation will come to dominate. 

One can also see that $\Gamma_\psi \gtrsim H$, showing that the fermions maintain a near-equilibrium distribution and that the inflaton's motion is adiabatic compared to the main microphysical processes in the thermal bath. In the bottom plot of Fig.~\ref{quartic_dynamics}, one can see that the temperature satisfies the conditions $gM\lesssim T\lesssim M$ for $M \simeq10^{15}$ GeV.

We note that the inflaton field value is much larger than the symmetry-breaking scale in this example, corresponding to a large relative phase between the two complex Higgs fields, although the inflationary energy scale $V_*^{1/4}\simeq 5\times 10^{15}$ GeV is comparable to $M$. Such large field values are natural in the context of chaotic inflation, where the Universe emerges from the pre-Planckian era with Planckian energy densities and a chaotic field distribution. The spatial regions that begin inflating already at this stage are those where the inflaton's potential energy is Planckian and for small self-couplings this corresponds to super-Planckian field values.

In this example, we find $n_s\simeq 0.964$ and $r\simeq 8\times 10^{-4}$ for thermal inflaton fluctuations, in agreement with the bounds set by the Planck satellite. We have explicitly checked that the Coleman-Weinberg corrections in Eq.~(\ref{finite_T_potential}) do not significantly change these observables nor the number of {\it e}-folds, although they become more relevant for larger values of $Q_*$. We will provide more details on this and other scalar potentials in a companion paper.

Our results show that there exist observationally consistent scenarios where inflation naturally occurs in a warm rather than supercooled regime, using simple interactions involving the inflaton and only four additional fields. A crucial issue in realizing warm inflation is to protect the flatness of the inflaton potential at a finite temperature, without suppressing dissipation. We have, for the first time, achieved this goal using symmetries that can also be employed to stabilize the Higgs boson mass. Our construction eliminates, in particular, the troublesome thermal corrections to the inflaton mass while still allowing for significant dissipative effects, and the fields coupled to the inflaton remain light despite the large field values generically required to sustain the slow roll dynamics for sufficiently long. These fields can also decay faster than expansion and thus keep a nearly thermal distribution, such that the dissipative process is adiabatic. We have, thus, evaded the major obstacles for realizing warm inflation in the high-temperature regime~\cite{YL} and, moreover, considering only a small number of light fields.

The presence of a thermal bath and dissipation during inflation can have a large impact on inflationary model building \cite{Berera:1999ws}. Dissipation alleviates the ``$\eta$ problem" that plagues supergravity or string inflation models and lowers the required field values in chaotic scenarios, thus significantly contributing towards constructing consistent effective field theory descriptions of inflation and their embedding within a fundamental theory. It also addresses the problem of initial conditions for inflation \cite{Berera:2000xz} and its ``graceful exit" into a radiation-dominated regime, eliminating the additional model dependence of a separate reheating period that may prove extremely hard to probe observationally. Moreover, in warm inflation important processes such as baryogenesis may occur during inflation and be probed using CMB data \cite{Berera:1996fm, Brandenberger:2003kc, BasteroGil:2011cx, Bastero-Gil:2014oga}. 

Realizing warm inflation in a simple model with only a few fields thus marks a significant step in constructing successful particle physics models of inflation where these features can be optimally developed.

\vspace{0.1cm} 
\begin{acknowledgments}

A.\,B.~is supported by STFC. M.\,B.\,-G.~is partially supported by ``Junta de Andaluc\'ia'' (FQM101) and the University of Granada (PP2015-03).  R.\,O.\,R.~is partially supported by  Conselho Nacional de Desenvolvimento Cient\'{\i}fico e Tecnol\'ogico - CNPq (Grant No.~303377/2013-5) and Funda\c{c}\~ao Carlos Chagas Filho de Amparo \`a Pesquisa do Estado do Rio de Janeiro - FAPERJ (Grant No.~E-26/201.424/2014). J.\,G.\,R. is supported by the FCT Grant No.~SFRH/BPD/85969/2012 and partially by the H2020-MSCA-RISE-2015 Grant No. StronGrHEP-690904, and by the CIDMA Project No.~UID/MAT/04106/2013.

 \vfill
\end{acknowledgments}


\begin{thebibliography}{9}
\bibitem{note1} This agrees with \cite{YL}, although with a somewhat smaller numerical factor due to improved approximations. This has, however, no significant impact on our conclusions.
\bibitem{note2} The interchange symmetry imposes $V(\phi)= V\left({\phi\over M} -{\pi\over4}\right)$, but this does not constrain the potential for $\phi\gg M$.
\end{thebibliography}
\end{document}